\documentstyle[12pt,lscape]{article}
\begin{document}
\title{\vspace*{-1.8cm}
Elastic and inelastic SU(3)-breaking final-state interactions
 in $B$ decays to
pseudoscalar mesons}
\author{
{P. \.Zenczykowski$^*$ and P. \L{}ach}\\
\\
{\em Dept. of Theoretical Physics},
{\em Institute of Nuclear Physics}\\
{Polish Academy of Sciences}\\
{\em Radzikowskiego 152,
31-342 Krak\'ow, Poland}\\
}
\maketitle
\begin{abstract}
We discuss all contributions from the Zweig-rule-satisfying  
SU(3)-breaking final state interactions (FSIs)
in the $B\to PP $ decays (neglecting charmed intermediate states),
where
$PP$=$\pi \pi$, $\pi K$, $K\bar{K}$,
$\pi \eta (\eta ')$, and $K \eta (\eta ')$. 
First, effects of
SU(3) breaking in rescattering through Pomeron exchange are studied.
Then, after making a plausible
assumption concerning the pattern of SU(3) breaking 
in non-Pomeron FSIs, we give general formulas for how
the latter modify short-distance (SD) amplitudes. In the SU(3) limit,
these formulas depend on
three effective
parameters 
 characterizing the strength of all non-Pomeron 
rescattering effects.
We point out that the experimental bounds on the $B \to K^+K^-$ 
branching ratio may limit the value of only one of these FSI parameters.
Thus, the smallness of the $B \to K^+K^-$ decay rate does not 
imply negligible rescattering effects in other decays.  Assuming
a vanishing value of this parameter, we perform various
fits to the available $B \to PP$ branching ratios. 
The fits determine the quark-diagram SD amplitudes, the 
two remaining FSI parameters
and the weak angle $\gamma $. 
While the set of all $B \to PP$ branching ratios is well described
 with $\gamma $ around its expected Standard Model (SM) value, 
 the fits permit
other values of $\gamma $  as well. For a couple of such good fits,
we predict asymmetries for the $B \to K \pi$, $\pi ^+ \eta (\eta ')$, 
$K^+ \eta (\eta ')$ decays as well as 
the values of the CP-violating parameters
$S_{\pi \pi}$ and $C_{\pi \pi}$ for the time-dependent
rate of $B^0(t) \to \pi ^+ \pi ^-$.
Apart from a problem with the recent $B^+ \to \pi ^+ \eta$ asymmetry
measurement,
comparison with the data seems to favour the values of $\gamma $ 
in accordance with SM expectations.

\end{abstract}
\noindent PACS numbers: 13.25.Hw, 11.30.Hv,12.15.Hh,11.80.Gw\\
$^*$ E-mail:
zenczyko@iblis.ifj.edu.pl
\newpage

\section{Introduction}
The majority of the analyses of CP-violating effects in $B$ decays assume
that the relevant amplitudes are given by short-distance (SD) expressions 
only. In particular, for $B$ decays into two pseudoscalar mesons
($B \to PP$),
any possible final state interactions (FSIs) are usually completely neglected. 
It is very difficult to assess if this neglect is justified or not.
Some authors have argued that such effects
should be negligible \cite{perturb,PQCD} since the $B$ mass is already quite
large.
In other papers it is stressed 
that the FSIs should be important and that any reliable analyses of $B$ decays
  must take these interactions  into account 
\cite{Wolf,FSI,Hou,Zen2001,otherFSI}. 
It has been suspected that the inelastic FSIs are particularly important 
\cite{Wolf,Zen2001}.
Unfortunately,
with our insufficient knowledge of the $PP$ interactions at $5.2~GeV$, there is
virtually no hope that the relevant
rescattering effects may be calculated reliably.

In order to overcome this obstacle,
in a recent paper \cite{LZ2002} we analysed an SU(3)-symmetric approach
with the built-in Zweig rule,
in which our ignorance as to the size of inelastic rescattering
was reduced to a set of only three {\em effective} (complex) parameters jointly
describing all inelastic final state interaction (IFSI) effects.
It was shown that the SU(3)-symmetric rescattering leads to a simple
redefinition of quark-diagram amplitudes, thus
permitting the use of a diagram
description in which, however, weak phases may enter in a modified way.
Furthermore, a simple estimate was made as to the size of error which could be
committed
while extracting the value
of the unitarity-triangle angle $\gamma$ when such modifications are not
taken into account.

In the present paper, we extend the general scheme of ref.\cite{LZ2002} and
introduce SU(3) breaking both in the elastic and in the inelastic
final state interactions.
The introduction of SU(3) breaking makes it reasonable to attempt
a detailed description of the data. When doing so, we
take into account all short-distance amplitudes usually
considered as the dominant ones (Section 2), and make certain assumptions 
as to the form
of FSIs and SU(3) breaking (Sections 3 and 4). 
In Section 4 we also discuss at some length the point 
that estimating the size of all rescattering effects
on the basis of the $B \to K\bar{K}$ data is significantly more difficult
than usually acknowledged.
Then, in Section 5, we perform fits to the experimental
branching ratios of the $B \to PP$ decays, and discuss their implications.
A brief summary appears in Section 6.

\section{Short-distance amplitudes}
Short-distance amplitudes may lead not only to the $PP$ states but also
to the general  $M_1M_2$  states, with $M_i$ representing various
heavy mesons. Consequently, the $PP$ pair observed in $B$ decay
may be produced in three ways:
it may not participate in any rescattering 
after being produced in a SD process, it may 
undergo elastic rescattering, and, finally, it may result from inelastic
rescattering of $M_1M_2$ into $PP$.
As discussed in \cite{LZ2002}, with the help of the unitarity condition,
contributions from other inelastic intermediate states (such as many-body
states $M_1M_2...M_n$)
may be always incorporated into the contribution from $M_1M_2$.

All SD amplitudes $B \to M_1M_2$ may be classified in the same way
as standard SD amplitudes $B \to PP$, ie.
$T,T'$ (tree), $C,C'$ (colour
suppressed), $P,P'$ (penguin), $E,E'$ (exchange), $A,A'$ (annihilation),
$PA,PA'$ (penguin annihilation), $S,S'$ (singlet penguin), 
$SS,SS'$ (double singlet penguin).  As usual, we denote
strangeness-conserving
$\Delta S =0$ (strangeness-violating $|\Delta S| =1$)
processes by unprimed (primed) amplitudes respectively.
Electroweak penguin 
contributions may be included via the replacements:
$T \to T+P^c_{EW}$, $P \to P-P^c_{EW}/3$,
$C \to C+P_{EW}$, $S \to S-P_{EW}/3$ \cite{elpenguins} (with analogous
expressions for the primed amplitudes).

The essential assumption of refs.\cite{Zen1,LZ2002} is that the tree,
penguin, etc. amplitudes
for the production of various $M_1M_2$ states are {\em proportional}
to the corresponding amplitudes for the production of the $PP$ pair.
One may argue that the relevant coefficients of proportionality
are approximately independent of the
 diagram type (tree, penguin, etc.) considered. The common remaining single
coefficient of proportionality may be absorbed into 
the rescattering amplitudes $M_1M_2 \to PP$, for
which the Zweig-rule is assumed.
Finally, 
the sum over all intermediate states $M_1M_2$ may be performed leading
to the appearance of only three effective complex parameters 
representing the relevant sums and
corresponding
to the presence of three Zweig-rule satisfying SU(3)-symmetric forms 
for $M_1M_2 \to PP$
(for more details, see \cite{Zen1,LZ2002}).

As a result of these simplifications, all contributions from various
short-distance $B \to M_1M_2$ amplitudes 
get expressed in terms of relevant standard
$B \to PP$ short-distance
amplitudes. Our whole approach to inelastic rescattering
depends therefore on standard
$T,P,...P',T',...$ etc. amplitudes (with appropriate weak phases) 
and on parameters effectively
describing the rescattering.
In order to simplify the discussion and study the effect of FSIs only,
we assume that the strong SD phases are negligible. (In ref.\cite{perturb}
these phases were estimated to be of the order of $10^o$, 
while in ref. \cite{GerardSmith2003} it is
argued that the FSI-uncorrected "bare" amplitudes 
do not contain any strong phases -
 see the comment after Eq.(16) therein). 
This assumption may be relaxed in future.

Some of the SD quark-diagram amplitudes are related. In an approach in which
FSIs break SU(3), one should incorporate
SU(3) breaking into the SD relationships as well. Therefore,
we assume that the tree SD amplitudes satisfy the following
relation \cite{CR2001}:
\begin{equation}
T'=\frac{V_{us}}{V_{ud}} \frac{f_K}{f_{\pi}} T \approx 0.276 ~T 
\end{equation}
Both tree amplitudes have the same (weak) phase: 
$T/|T|=T'/|T'|=e^{i \gamma }$.

The penguin SD amplitudes are dominated by the $t$ quark, so that 
the weak phase factor is $e^{-i \beta}$ for $P$ and $\pm 1$ for $P'$
(ie. $P'=\pm |P'|$). 
We use the estimate \cite{CR2001}
\begin{equation}
\label{penguins}
P=-e^{-i \beta}\left | \frac{V_{td}}{V_{ts}}\right | P'
\approx -0.176~ e^{-i \beta }P'.
\end{equation} 
In the fits of Section 5, we accept $\beta = 24^o$,
which is in agreement with the world average \cite{beta}
$\sin 2\beta = 0.734 \pm 0.054$. 
We accept (as it is usually done) 
that the value of the penguin SD amplitudes
does not depend on the flavour of the quark-antiquark pair created to produce 
the $M_1M_2$ state.
For example, standard SD contributions from penguin $P$ 
in $B^0_d \to \pi^+\pi^-$ (or $\pi ^0 \pi^0$), and in $B^+ \to
K^+\bar{K}^0$ are given by SU(3) considerations only, despite the fact
that in these two
processes the produced quark-antiquark pairs are of different flavours.


We accept the relations between
the tree and the colour-suppressed amplitudes given by the SD estimates:
\begin{equation}
C = \xi T
\end{equation}
and
\begin{equation}
\label{cprime}
C'=T'(\xi -(1+\xi) \delta _{EW} e^{-i\gamma})
\end{equation}
where we take $\xi =\frac{C_1+\zeta C_2}{C_2+\zeta C_1}\approx 0.17 $,
assuming $\zeta \approx 0.42$, ie. midway between $1/N_c$ and the value of
$0.5$ suggested by experiment, and using $C_1\approx -0.31$ and $C_2 \approx 
1.14$ \cite{BBurasL}.
The contribution from the electroweak penguin $P'_{EW}$ 
has been included in Eq.(\ref{cprime}),
with $\delta _{EW}\approx +0.65$ \cite{EWP} (other
electroweak penguins are neglected).

The last independent
SD amplitude considered here is the singlet penguin
amplitude $S'$, whose weak phase is $0$ 
(data requires that this amplitude be sizable \cite{DGR97PRL79,CR2001}).
Thus, the SD amplitudes and our whole approach depend on four SD parameters: 
$|T|$, $P'$, $S'$, and the weak phase $\gamma $.
The remaining SD amplitudes ($E,E',S,PA,...$) are assumed
to be negligible.

\section{SU(3)-breaking in Pomeron-exchange-induced rescattering}

If we gather all SD amplitudes $B\to PP$ (as well as those of $B \to M_1M_2$) 
into vector ${\bf w}$, and
accept that FSIs cannot modify the probabilities of the original SD
weak decays, it follows that vector ${\bf W}$ representing the set
of all FSI-corrected amplitudes is related to ${\bf w}$ through 
\cite{Zen2001,Zen1}:
\begin{equation}
\label{S12}
{\bf W}={\bf S}^{1/2}{\bf w},
\end{equation}
(in the one-channel case, Eq.(\ref{S12}) reduces to the Watson's theorem
\cite{Watson}).

Let us consider now elastic $PP$ rescattering only (ie. with 
${\bf w}$ restricted to its part corresponding
to $B \to PP$ processes, and similarly for ${\bf W}$).
For high energies this rescattering is approximately 
independent of energy.   We shall use Regge terminology and call
this energy-independent term a
Pomeron-induced contribution.
Since Pomeron exchange is known to be substantial, 
the $B \to PP$ amplitudes at $s =m^2_B$ should be corrected for
Pomeron-induced rescattering. 
Treating Pomeron-induced FSIs as a small correction
to the SD expressions for $B \to PP$ amplitudes corresponds to 
expanding ${\bf S}^{1/2}\equiv ({\bf 1}+i{\bf T})^{1/2}=
({\bf 1}+2i{\bf A})^{1/2}={\bf 1}+i{\bf A}+...$ 
and keeping terms linear in
${\bf A}$ only. Thus, one gets \cite{Zen2001}:
\begin{equation}
\label{jedenPLUSiA}
{\bf W}\approx ({\bf 1}+i {\bf A}){\bf w}
\end{equation}
Because the amplitudes for Pomeron exchange
are predominantly
imaginary, we have
\begin{equation}
{\bf A}=i {\bf a}
\end{equation}
with real ${\bf a}$.
In the SU(3)-symmetric world, all elements of ${\bf a}$ are identical,
and their common value is $a \approx 0.16 $
 (cf \cite{Wolf} and Eqs.(10,17) in \cite{Zen2001}). 
Consequently, Pomeron-induced rescattering  rescales
all SD amplitudes in the same way:
\begin{equation}
{\bf W}=(1-a){\bf w}.
\end{equation}
It is only when SU(3) is broken that the rescaling is different for
different decay channels, and deviations from the standard SD form could
 be observed in principle.

When SU(3) is broken, the values of $a$ differ
for different final channels $P_1P_2$. 
In a simple model for Pomeron used in
\cite{Wolf,Lach}, they  are given by
\begin{equation}
\label{abroken}
a(P_1P_2)=\frac{1}{16 \pi} \frac{\beta_{P_1}\beta_{P_2}}{b_{P_1}+b_{P_2}}
\end{equation}
with the values of  $\beta_{\pi }$, $\beta_{K}$
(meson-Pomeron couplings) and 
  $b_{\pi }$, $b_{K}$ (slope coefficients for
 the relevant
couplings) extracted from data
on $\pi p$ and $K p$ scattering.
In the following we will use the averages of values given in
\cite{Wolf,Lach}, ie.:
\begin{eqnarray}
\beta_{\pi}&=& 3.47~ \sqrt{{\rm mb}}\nonumber \\
\beta_K&=& 2.78~ \sqrt{{\rm mb}}\nonumber \\
b_{\pi}&=&1.93 ~{\rm GeV}^{-2}\nonumber \\
b_K&=&0.9 ~{\rm GeV}^{-2}
\end{eqnarray}
In order to estimate
 $\beta _{\eta} $, $\beta_{\eta '}$ and $b_{\eta}$, $ b_{\eta '}$,
 we assume perfect mixing for $\eta, \eta '$ 
 (ie. $\eta = (u\bar{u}+d\bar{d}-s\bar{s})/\sqrt{3}$, and
  $\eta '= (u\bar{u}+d\bar{d}+2s\bar{s})/\sqrt{6}$) 
  corresponding to the octet-singlet mixing angle of $\theta = -19.5^o$, 
  (see eg.
  \cite{Lipkin81,Chau91D43,GR96D53,DGR97PRL79};
  for a different approach to $\eta - \eta '$ mixing in $B \to K 
  \eta^{(}\phantom{}'^{)}$
  decays see \cite{BN2003}),
  and
 derive \cite{Lach}:
 \begin{eqnarray}
 \beta _{\eta}&=& (\beta_{\pi}+2 \beta_K)/3 \approx 3.01 ~\sqrt{{\rm mb}}
 \nonumber \\
 \beta_{\eta '}&=&(-\beta _{\pi}+4 \beta _K)/3 \approx 2.55~\sqrt{{mb}}
 \nonumber \\
 b_{\eta}&=&(b_{\pi}+ 2b_K)/3 \approx 1.24~{{\rm GeV}}^{-2}\nonumber \\
 b_{\eta '}&=& (-b_{\pi}+4 b_K)/3 \approx 0.56~{{\rm GeV}}^{-2}
 \end{eqnarray}
 Note that for the 
 $K \eta'$ channel the denominator in Eq.(\ref{abroken})
 is particularly small. In this channel 
 the Pomeron-exchange-induced correction
is therefore relatively large which may possibly affect the extraction of the
short-distance $S'$ amplitude from the data.

The resulting pattern
of SD amplitudes corrected for Pomeron-induced
rescattering differs from standard SD expressions 
by departures from SU(3) only. 
Consequently, we introduce SU(3)-symmetric rescaled amplitudes $\bar{T}$,  
$\bar{T}'$, $\bar{P}$, $\bar{P}'$, ..etc., defined as
\begin{eqnarray}
\bar{T}^{(')}&=&T^{(')}(1-a(\pi \pi))\nonumber \\
\bar{P}^{(')}&=&P^{(')}(1-a(\pi \pi))\\
&...&, \nonumber
\end{eqnarray}
and the SU(3)-breaking corrections 
$K(P_1P_2)= (a(\pi \pi)-a(P_1P_2))/(1 - a(\pi \pi))$.
The complete set of SD amplitudes corrected 
for SU(3)-breaking Pomeron-exchange-induced
 rescattering
is gathered in Table 1.

\begin{table}
\caption{SD amplitudes for decays $B^+,B^0_d \to P_1P_2$
corrected for SU(3)-breaking Pomeron-exchange-induced rescattering
}
\label{elastic}
\begin{center}
\begin{footnotesize}
\begin{tabular}{ccc}
\hline
Decay  & rescaled and corrected SD &  \\
\hline
$B^+ \to$ $ \pi ^+ \pi ^0$       & 
$-\frac{1}{\sqrt{2}}(\bar{T}+\bar{C})$ &  \\
\phantom{$B^+ \to$} $K^+\bar{K}^0 $&  $ -\bar{P} (1 +K(KK))$& 
\\
\phantom{$B^+ \to$} $\pi ^+\eta $&
$-\frac{1}{\sqrt{3}}(\bar{T}+\bar{C}+2\bar{P})(1+K(\pi \eta))$&
\\
\phantom{$B^+ \to$} $\pi ^+\eta '$
&$-\frac{1}{\sqrt{6}}(\bar{T}+\bar{C}+2\bar{P})(1+K(\pi \eta '))$&\\
\hline
$B^0_d \to$ $\pi ^+ \pi ^-$ &$ -(\bar{T}+\bar{P}) $&   \\
\phantom{$B^0_d \to$} $\pi ^0 \pi ^0$ 
&$-\frac{1}{\sqrt{2}}(\bar{C}-\bar{P})$&\\
\phantom{$B^0_d \to$} $K^+K^-$ & $0$ &\\
\phantom{$B^0_d \to$} $K^0 \bar{K}^0$ &$-\bar{P}(1+K(KK))$&\\
\hline
$B^+ \to $ $\pi ^+ K^0$& $-\bar{P}'(1+K(\pi K))$&\\
\phantom{$B^+ \to$} $\pi ^0 K^+$
&$\frac{1}{\sqrt{2}}(\bar{T}'+\bar{C}'+\bar{P}')(1+K(\pi K))$&\\
\phantom{$B^+ \to$} $\eta K^+$
&$\frac{1}{\sqrt{3}}(\bar{T}'+\bar{C}'+\bar{S}')(1+K(\eta K))$&\\
\phantom{$B^+ \to$} $\eta ' K^+$& 
$\frac{1}{\sqrt{6}}(\bar{T}'+\bar{C}'+3\bar{P}'+4\bar{S}')(1+K(\eta ' K))$&\\
\hline
$B^0_d \to$ $\pi ^- K^+$& $(\bar{T}'+\bar{P}')(1+K(\pi K))$&\\
\phantom{$B^0_d \to$} $\pi ^0 K^0$ 
&$\frac{1}{\sqrt{2}}(\bar{C}'-\bar{P}')(1+K(\pi K))$&\\
\phantom{$B^0_d \to$} $\eta K^0$ &
$\frac{1}{\sqrt{3}}(\bar{C}'+\bar{S}')(1+K(\eta K))$&\\
\phantom{$B^0_d \to$} $\eta ' K^0$ &
$\frac{1}{\sqrt{6}}(\bar{C}'+3\bar{P}'+4\bar{S}')(1+K(\eta ' K))$&\\
\hline
\end{tabular}
\end{footnotesize}
\end{center}
\end{table}

\section{Inelastic SU(3)-breaking FSI with Zweig rule}

Analysis of inelastic SU(3)-breaking effects follows the approach
of \cite{LZ2002}. 
As in ref. \cite{LZ2002}, in the present paper we do not consider
contributions from intermediate charmed states
(thus neglecting the long-distance "charming penguins"). Since they may
be important 
 \cite{Ciuchini,BurasSilvestrini,Pham,Rosner2001,GerardSmith2003},
their analysis 
merits further work.
The most general
Zweig-rule-satisfying rescattering $M_1M_2 \to P_1P_2$ 
is described by two types of connected
diagrams:
the "uncrossed" diagrams of Fig.1($u$), and the "crossed" diagrams of 
Fig.1($c$).
By virtue of Bose statistics, the final $P_1P_2$ pair must be in 
an overall symmetric state.
Our definition of inelastic rescattering includes a non-Pomeron contribution 
from $P_1P_2 \to P_1P_2$ transitions, which - together with the
Pomeron-exchange-induced part of these 
transitions - are usually classified as elastic.

\subsection{SU(3)-invariant rescattering amplitudes}
In the SU(3) case, the requirement of Bose statistics for $P_1P_2$
means that there are two 
types of uncrossed $M_1M_2 \to P_1P_2$
amplitudes, ie. 
(using a particle symbol for the
corresponding SU(3) matrix):
\begin{equation}
\label{uncrosseds}
{\rm Tr}(\{M_1^{\dagger},M_2^{\dagger}\}\{P_1,P_2\})~u_+
\end{equation} 
and
\begin{equation}
\label{uncrosseda}
{\rm Tr}([M_1^{\dagger},M_2^{\dagger}]\{P_1,P_2\})~u_-
\end{equation}
where the requirement in question
 is reflected by the presence of the anticommutator
$\{P_1,P_2\}$ of meson matrices,
and $u_{\pm}$ denote the strength of rescattering amplitudes.
Eqs.(\ref{uncrosseds},\ref{uncrosseda}) incorporate
nonet symmetry for both intermediate and final mesons.
As explained in \cite{LZ2002}, invariance of  strong
interactions under charge conjugation demands that
mesons $M_1$ and $M_2$ belong to multiplets of the same (opposite)
C-parities for
the first (second) amplitude above.

For the crossed diagrams, the requirement of $P_1 \rightleftharpoons P_2$
symmetry admits one combination only \cite{LZ2002}:
\begin{equation}
\label{crossed}
{\rm Tr}(M_1^{\dagger}P_1M_2^{\dagger}P_2
+M_1^{\dagger}P_2M_2^{\dagger}P_1)~c
\end{equation}
where $c$ denotes amplitude strength. 
This combination is symmetric under $M_1\rightleftharpoons M_2$ as well.
Consequently, it
is charge-conjugation invariant if $M_1$ and $M_2$ have 
C-parities of the same sign.

\subsection{Modifications due to SU(3) breaking}
We will incorporate SU(3) breaking into the FSI amplitudes of 
Eqs.(\ref{uncrosseds}, \ref{uncrosseda}, \ref{crossed})
in the simplest possible way. 
First let us consider  $u$-type diagrams (Fig.1(u)).
In these diagrams one quark (or antiquark) from meson $M_{1}$ 
ends up in the final pseudoscalar meson, while the other one annihilates
an antiquark (quark) from meson $M_{2}$. It is well known
that such quark-antiquark annihilations  
are suppressed when the relevant $q\bar{q}$ pair has
high energy, 
and that they are suppressed
even more strongly for the $s\bar{s}$ pair.
 In the Regge language, the first statement corresponds to
 meson exchanges being suppressed
at high energies, the latter - to
the fact that intercepts
of Regge trajectories for mesons containing strange quarks lie below
those for mesons composed of $u,d,\bar{u},\bar{d}$ only.
The additional suppression of $s\bar{s}$ annihilation with
respect to that of $u\bar{u}$ (or $d\bar{d}$) depends
on the energy of the $q\bar{q}$ pair. Since we want to analyse 
the main effect of SU(3) breaking only,   we 
assume that an exchange of a strange (anti)quark between 
mesons $M_1$ and $M_2$ (or between $P_1$ and $P_2$) is 
suppressed by the same factor ($\epsilon $) for all intermediate
states.
On the other hand, the amplitudes for the uncrossed diagrams
in which strange (anti)quarks from mesons $M_1$ end up in final
pseudoscalar mesons (ie. are not annihilated) are not suppressed 
by SU(3)-breaking
effects.

The relevant $u$-type amplitudes may be then calculated from the
appropriate generalizations of Eqs.(\ref{uncrosseds},\ref{uncrosseda}).
For the contribution from mesons $M_1$ and $M_2$ of the
same charge-conjugation parities ($C(M_1)C(M_2)=+1$) we have, for example:
\begin{eqnarray}
&\frac{1}{2}{\rm Tr}((M_1^{\dagger }I_{\epsilon } M_2^{\dagger }
+M_2^{\dagger }I_{\epsilon } M_1^{\dagger })
(P_1 I_{\epsilon }P_2+P_2 I_{\epsilon }P_1))~u_+&\nonumber \\
\label{uncrossedsbreak}
+&\frac{1}{2}{\rm Tr}((M_1^{\dagger T} I_{\epsilon } M_2^{\dagger T}
+M_2^{\dagger T} I_{\epsilon } M_1^{\dagger T})
(P_1^T I_{\epsilon }P_2^T+P_2^T I_{\epsilon }P_1^T))~u_+&
\end{eqnarray} 
where
\begin{equation}
\label{Ieps}
I_{\epsilon }=\left[ \begin{array}{ccc}
1&0&0\\
0&1&0\\
0&0& \epsilon 
\end{array}\right].
\end{equation}
In Eq.(\ref{uncrossedsbreak}) we divided the whole contribution into
two parts, depending on whether it is the strange quark or antiquark
from (say) $M_1$ which is annihilated.
Contributions from the $C(M_1)C(M_2)=-1$ states may be calculated
in a similar way (one has to remember that the negative sign between
$M_1 I_{\epsilon } M_2$ and $M_2 I_{\epsilon } M_1$ is cancelled by the
negative sign in the (antisymmetric) 
wave function of $C(M_1)C(M_2)=-1$ two-meson states).

Since SU(3) is to be broken, the choice of 
definite SU(3) representations for the 
intermediate $M_1M_2$ states is not appropriate. Admitting 
the linear combinations
of ${\bf 27}$, ${\bf 8_s}$, ${\bf 8_{\{81\}}}$, 
${\bf 1_{\{88\}}}$, ${\bf 1_{\{11\}}}$, and ${\bf 8_a}$ 
(considered in \cite{LZ2002})
is not sufficient either, since for broken SU(3)
the complete set of $M_1M_2$ intermediate states contains admixtures
from other SU(3) representations. 
If all
the $C(M_1)C(M_2)=\pm 1$ intermediate states are to be taken into account
properly, one may first list all states of definite charge, strangeness
and isospin, and composed of two mesons of definite type, ie. with flavour
quantum numbers of $\pi K$ {\em or}
$\eta K$ {\em or}... . These states may be ordered
(in the sense that $\pi K$ is different from $K \pi$) or,
alternatively, their symmetric or antisymmetric combinations
(under $\pi \leftrightarrow K$ etc. interchanges) may be formed. 
Then, SD decay amplitudes into these states have to be evaluated.
Finally, the sum over the contributions from 
all such states has to be carried out.

We have performed all the necessary calculations with the 
result that the sum over {\em all} $C(M_1)C(M_2)=\pm 1$ 
intermediate states leads to the formulas given in the second column of
Table 2, where
\begin{eqnarray}
\bar{u}&=&u~\frac{1}{1-a(\pi \pi)}=
\frac{u_++u_-}{2}\frac{1}{1-a(\pi \pi)}\nonumber \\
\label{barFSI}
\bar{d}&=&d~\frac{1}{1-a(\pi \pi)}=(u_+-u_-)\frac{1}{1-a(\pi \pi)}
\end{eqnarray}
and
\begin{eqnarray}
\Delta  &=&((2+\epsilon )\bar{P}+\bar{T})~\bar{d}\nonumber \\
\label{Deltadefinition}
\Delta '&=& ((2+\epsilon )\bar{P}'+\bar{T}'+\epsilon \bar{S}')~\bar{d}.
\end{eqnarray}
Thus, the inelastic corrections are given in terms of the products of the SD
amplitudes and the FSI parameters (here: $u$, $d$). For example, there may be a
contribution proportional to $Td$. Since we finally express all formulas in
terms of the amplitudes modified for Pomeron-induced rescattering (eg. in terms
of $\bar{T}=T(1-a(\pi \pi))$ etc.), 
in Eq.(\ref{barFSI}) we introduced the rescaled FSI
parameters $\bar{u}$ and $\bar{d}$ so that eg. $Td=\bar{T}\bar{d}$.
For completeness, in Table 2 we give formulas for
the $B^0_s$ decays as well.

\begin{table}
\caption{Inelastic SU(3)-breaking rescattering contributions: $\Delta
\equiv((2+\epsilon )\bar{P}+\bar{T})~\bar{d}$; 
$\Delta ' \equiv ((2+\epsilon )\bar{P}'+\bar{T}'+\epsilon \bar{S}')~\bar{d}$}
\label{inelastic}
\begin{center}
\begin{footnotesize}
\begin{tabular}{ccc}
\hline
Decay  & uncrossed FSI diagrams & crossed FSI diagrams \\
\hline
$B^+ \to$ $ \pi ^+ \pi ^0$       & 
$0$ &  $-\frac{1}{\sqrt{2}}2\bar{c}(\bar{T}+\bar{C})$ \\
\phantom{$B^+ \to$} $K^+\bar{K}^0 $& 
$ -\epsilon (\Delta + 2 \bar{u} \bar{C})$ & 0
\\
\phantom{$B^+ \to$} $\pi ^+\eta $&
$-\frac{2}{\sqrt{3}}(\Delta+2 \bar{u}\bar{C})$& 
$-\frac{1}{\sqrt{3}} 2\bar{c}(\bar{T}+\bar{C}+\bar{P}(2-\epsilon))$
\\
\phantom{$B^+ \to$} $\pi ^+\eta '$
&$-\frac{2}{\sqrt{6}}(\Delta+2 \bar{u}\bar{C})$&
$-\frac{1}{\sqrt{6}} 2\bar{c}(\bar{T}+\bar{C}+2\bar{P}(1+\epsilon))$
\\
\hline
$B^0_d \to$ $\pi ^+ \pi ^-$ &$ -(\Delta + 2\bar{u}(\bar{T}+2\bar{P})) $&  
$-2\bar{c} \bar{C}$ \\
\phantom{$B^0_d \to$} $\pi ^0 \pi ^0$ 
&$ \frac{1}{\sqrt{2}}(\Delta + 2\bar{u}(\bar{T}+2\bar{P})) $&
$-\frac{1}{\sqrt{2}}2\bar{c} \bar{T}$\\
\phantom{$B^0_d \to$} $K^+K^-$ & $2\bar{u}(\epsilon \bar{T} +
(1+\epsilon )\bar{P})$ & $0$\\
\phantom{$B^0_d \to$} $K^0 \bar{K}^0$ &$-\epsilon \Delta -2\bar{u}
(1+\epsilon )\bar{P} $& $0$\\
\phantom{$B^0_d \to$} $\pi ^0 \eta$ &
$-\frac{2}{\sqrt{6}} (\Delta -2\bar{u} \bar{T})$&
$-\frac{2}{\sqrt{6}}(2-\epsilon )\bar{c} \bar{P}$\\
\phantom{$B^0_d \to$} $\pi ^0 \eta '$ &
$-\frac{1}{\sqrt{3}}(\Delta - 2 \bar{u} \bar{T})$&
$-\frac{1+\epsilon }{\sqrt{3}}2\bar{c}\bar{P}$\\
\hline
$B^0_s \to$ $\pi ^+ K^-$&$-\Delta $& $-(1+\epsilon ) \bar{c} \bar{C}$\\
\phantom{$B^0_s \to$} $\pi ^0 \bar{K}^0$&$\frac{1}{\sqrt{2}}\Delta $
&$-\frac{1}{\sqrt{2}}(1+\epsilon ) \bar{c}\bar{T}$\\
\phantom{$B^0_s \to$} $\eta \bar{K}^0$&
$-\frac{1-\epsilon }{\sqrt{3}}\Delta $& 
$-\frac{1+\epsilon}{\sqrt{3}}\bar{c}((2-\epsilon )\bar{P} +\bar{T})$\\
\phantom{$B^0_s \to$} $\eta ' \bar{K}^0$&
$-\frac{1+2 \epsilon }{\sqrt{6}}\Delta $&
$-\frac{1+\epsilon}{\sqrt{6}}\bar{c} (2(1+\epsilon )\bar{P}+\bar{T})$\\
\hline
\hline
$B^+ \to $ $\pi ^+ K^0$& $-\Delta ' - 2\bar{u} (\bar{C}'+\bar{S}')$&
$-\bar{c}(1+\epsilon )\bar{S}'$\\
\phantom{$B^+ \to$} $\pi ^0 K^+$
&$\frac{1}{\sqrt{2}}(\Delta ' + 2\bar{u} (\bar{C}'+\bar{S}'))$
&$\frac{1}{\sqrt{2}}\bar{c}(1+\epsilon )(\bar{T}'+\bar{C}'+\bar{S}')$\\
\phantom{$B^+ \to$} $\eta K^+$
&$\frac{1}{\sqrt{3}}(1-\epsilon )(\Delta ' + 2\bar{u} (\bar{C}'+\bar{S}'))$
&$\frac{1}{\sqrt{3}}\bar{c}(1+\epsilon )
(\bar{T}'+\bar{C}'+\bar{P}'(2-\epsilon)+\bar{S}'(1-\epsilon )$\\
\phantom{$B^+ \to$} $\eta ' K^+$& 
$\frac{1+2\epsilon}{\sqrt{6}}(\Delta ' + 2\bar{u} (\bar{C}'+\bar{S}'))$
&$\frac{1}{\sqrt{6}}\bar{c}(1+\epsilon
)(\bar{T}'+\bar{C}'+2(1+\epsilon)\bar{P}'+(1+2\epsilon )\bar{S}')$\\
\hline
$B^0_d \to$ $\pi ^- K^+$& $\Delta ' + 2 \bar{u} \bar{S}'$
&$\bar{c} (1+\epsilon )(\bar{C}'+\bar{S}')$\\
\phantom{$B^0_d \to$} $\pi ^0 K^0$ 
&$-\frac{1}{\sqrt{2}}(\Delta ' + 2 \bar{u} \bar{S}')$
&$\frac{1}{\sqrt{2}}\bar{c}(1+\epsilon )(\bar{T}'-\bar{S}')$\\
\phantom{$B^0_d \to$} $\eta K^0$ &
$\frac{1}{\sqrt{3}}(1-\epsilon )(\Delta ' + 2 \bar{u} \bar{S}')$
&$\frac{1}{\sqrt{3}}\bar{c}(1+\epsilon )
(\bar{T}'+(2-\epsilon )\bar{P}'+(1-\epsilon )\bar{S}')$\\
\phantom{$B^0_d \to$} $\eta ' K^0$ &
$\frac{1+2\epsilon}{\sqrt{6}}(\Delta ' + 2 \bar{u} \bar{S}')$
&$\frac{1}{\sqrt{6}}\bar{c}(1+\epsilon )
(\bar{T}'+2(1+\epsilon )\bar{P}'+(1+2 \epsilon )\bar{S}')$\\
\hline
$B^0_s \to$ $\pi ^+ \pi ^-$&
$-2 \epsilon \bar{u}(2\bar{P}'+\bar{T}')$&$0$\\
\phantom{$B^0_s \to$} $\pi ^0 \pi ^0$&
$\sqrt{2} \epsilon \bar{u} (2\bar{P}'+\bar{T}') $&
$0$\\
\phantom{$B^0_s \to$} $K^+K^-$&
$\Delta ' +2 \epsilon \bar{u} ((1+\epsilon )\bar{P}'+\epsilon \bar{T}'
+\bar{S}') $&
$2 \epsilon \bar{c} (\bar{C}'+\bar{S}')$\\
\phantom{$B^0_s \to$} $K^0\bar{K}^0$&
$-\Delta ' -2 \epsilon \bar{u} ((1+\epsilon )\bar{P}'+\bar{S}')$&
$-2 \epsilon \bar{c} \bar{S}'$\\
\phantom{$B^0_s \to$} $\pi ^0 \eta $&
$\frac{4}{\sqrt{6}}\epsilon \bar{u} \bar{T}'$&
$-\frac{2}{\sqrt{6}}\epsilon \bar{c} \bar{T}'$\\
\phantom{$B^0_s \to$} $\pi ^0 \eta '$&
$\frac{2}{\sqrt{3}}\epsilon \bar{u} \bar{T}'$&
$\frac{2}{\sqrt{3}}\epsilon \bar{c} \bar{T}'$\\
\hline
\hline
\end{tabular}
\end{footnotesize}
\end{center}
\end{table}

We incorporate
SU(3) breaking  into the $c$-type amplitudes in
a completely analogous fashion. Namely, 
we assume that strange (anti)quark interchanges are suppressed
by factor $\epsilon $ (in general, this factor may be different from
that used for $u$-type diagrams).
The relevant $c$-type amplitudes may be then calculated from
an appropriate generalization of Eq.(\ref{crossed}). As pointed out in
\cite{LZ2002},  charge conjugation
invariance of strong interactions requires that only symmetric $M_1M_2$ 
states contribute. For broken SU(3), Eq.(\ref{crossed}) is replaced by
\begin{eqnarray}
&\frac{1}{2}{\rm Tr}
(M_1^{\dagger}I_{\epsilon }P_1M_2^{\dagger}I_{\epsilon }P_2
+M_1^{\dagger}I_{\epsilon }P_2M_2^{\dagger}I_{\epsilon }P_1)~c&
\nonumber \\
+&\frac{1}{2}{\rm Tr}(M_1^{\dagger T}I_{\epsilon }P_1^T 
M_2^{\dagger ^T}I_{\epsilon }P_2^T
+M_1^{\dagger T}I_{\epsilon }P_2^T M_2^{\dagger T}I_{\epsilon }P_1^T)~c&
\end{eqnarray}
As
in Eq.(\ref{uncrossedsbreak}), we divided the whole contribution into
two parts depending on whether it is the strange quark or antiquark
from (say) $M_1$ which is exchanged.
Using the above expression and the expressions for the SD amplitudes, and
summing over all the intermediate states,
one obtains the corrections
 induced by the $c$-type IFSIs.  They are listed
  in the third column of
Table 2, where
\begin{equation}
\bar{c}=c/(1-a(\pi \pi))
\end{equation}
In the limit of $\epsilon \to 1$, all formulas of Table 2 reduce to those
given in \cite{LZ2002}, while for
$\epsilon = 0$ SU(3) is maximally broken.

\subsection{Structure of full FSIs}
For small inelastic final-state interactions, 
Eqs.(\ref{S12},\ref{jedenPLUSiA}) 
suggest the following approximation of all FSI effects:
\begin{equation}
\label{fullFSI}
{\bf W}\approx {\bf w}- {\bf a}{\bf w} + i {\bf \Delta W_{inel}} 
\end{equation}
where the three terms correspond to the contributions 
from the unmodified SD amplitudes, the Pomeron-exchange-induced corrections, 
and the
inelastic FSI corrections (including the $P_1P_2 \to P_1P_2$ elastic
transitions not mediated by Pomeron) respectively. Here
 ${\bf \Delta W_{inel}}$ (proportional to 
 $\sum _{M_1M_2}{\bf T}|M_1M_2\rangle\langle M_1M_2| {\bf w}$ ) 
 is given by expressions for the inelastic
FSIs gathered in Table 2. 
For negligible strong SD phases, it is the third term in Eq.(\ref{fullFSI}) 
which allows
the existence of direct CP violation effects.
This term provides a specific prescription for how strong
phases are generated by quark interchanges
between outgoing mesons. In other words, the pattern of FSI phases in all
$B \to PP $ decays
is governed by three (in general complex) 
parameters $\bar{d}$, $\bar{u}$, $\bar{c}$
corresponding to different flavour-flow
 rescattering topologies and by the value of the SU(3)-breaking parameter(s)
 $\epsilon$.

\subsection{Size of rescattering effects and ${\bf B \to KK} $ decays}
The Pomeron-induced FSIs and a
contribution from non-Pomeron-mediated transitions $P_1P_2 \to P_1P_2$
together comprise elastic rescattering.
The non-Pomeron contributions to
elastic transitions (eg. quark-line exchange diagrams for
 $\pi ^+ \pi ^- \to \pi ^+ \pi ^-$)
 should be treated alongside
symmetry-related contributions 
(ie. $\pi ^+ \pi ^- \to \pi ^o \pi ^o$ or $\pi ^+ \pi ^- \to K^+ K^-$ etc.),
 as they all have common origin.
 For the SU(3) case
all such "quasi-elastic" $P'_1P'_2 \to P_1P_2$ transitions 
were estimated in the Regge approach \cite{Zen1999}. The resulting
differences between strong phases 
in the singlet, octet, and 27-plet $PP$ channels
(see also \cite{Lach})
 vanish at high energy, while at the $B$-meson mass 
they turn out to be nonnegligible yet small, of the order of $10^o$. 
Consequently, inclusion of full elastic FSIs should not lead
to a significant change in the quality of data description 
(see also the fits of the next Section).

As for the inelastic rescattering, Table 2 provides the basis for the 
relevant discussion.

If FSIs satisfy SU(3) (ie. if $\epsilon =1$), all the
$\Delta$ and $\Delta '$ terms in Table 2
may be absorbed into the
new redefined penguins~\cite{LZ2002} (compare Eq.(\ref{fullFSI})):
\begin{eqnarray}
\tilde{P}&=&\bar{P}+ i\Delta\nonumber \\
\label{tildedef}
\tilde{P}'&=&\bar{P}'+i\Delta '
\end{eqnarray}
With our assumptions of SU(3)-symmetric
SD penguin amplitudes (cf. comment after Eq.(\ref{penguins})),
such a redefinition is possible only
if $\epsilon =1$  
(compare the relevant $\Delta $-dependent corrections 
to $B^+ \to K^+\bar{K}^0$ and 
$B^0_d \to \pi ^+ \pi^-$ in Table 2).
As can
be seen from the presence of the SD tree
amplitudes in the redefined penguins in
Eq.(\ref{tildedef}) (cf. $\Delta = (3\bar{P}+\bar{T})\bar{d}$),
parameter $\bar{d}$ is related
to the size of the long-distance ($u$-quark-loop) penguin.
Formulas (\ref{tildedef}) indicate
that 
contributions from penguin topologies with internal $u$-quark loops
cannot lead to significant modification of total amplitudes -
all such effects are consistently absorbed everywhere into new redefined
penguins $\tilde{P}$, $\tilde{P}'$. The only change is in the
phase factors since $\Delta $'s include terms depending on $\gamma $.
In general this leads to nonzero asymmetries, and should
affect the determination of $\gamma $, as the (effective) 
amplitudes will now interfere in a different way.

In some papers it was argued that the $B \to K^+K^-$ decays
could provide an estimate of the size of rescattering effects.
Note, however, that this decay amplitude depends on $u$
only.
The $B \to K^+K^-$ branching ratio is independent of $\bar{d}$ and
$\bar{c}$, and, consequently, the size of long-distance penguins
is not restricted by $B \to K^+ K^-$.
This means that
this branching ratio
is not such a good place to estimate the "typical" size of FSI effects as
it has been thought so far.

It is also sometimes said that the size of rescattering effects 
may be gleamed from the
$B^+ \to K^+\bar{K}^0$ decay which is
related to the $B^+ \to \pi^+ K^0$ decay by an interchange
of all down and strange quarks
\cite{Fleischerresc}. Here the standard argument assumes 
U-spin flavour symmetry of strong interactions.
When SU(3) is broken,
 a look at Table 2
and Eqs. (\ref{Deltadefinition},\ref{tildedef}) 
shows that the conclusions from the comparison of
$B^+ \to K^+\bar{K}^0$ and $B^+ \to \pi^+ K^0$ decays
{\em cannot} be obtained in such a simple way as originally thought.
Namely, with the contribution from $\bar{u}$-generated
FSI effects being bounded by the smallness of the $B^0_d \to K^+K^-$
branching ratios, the FSI effects in $B^+ \to K^+\bar{K}^0$ are 
proportional to term $\epsilon \Delta $. However, on the basis of Regge ideas 
and our knowledge of high-energy multiparticle
production processes in which $K\bar{K}$ pairs are rarely produced, one
expects that $\epsilon $ is small.
(The assumption of negligible $\epsilon $ seems to be
 corroborated by the $\epsilon $-dependence of our fits 
below.)
Consequently, the rescattering term in 
$B^+ \to K^+\bar{K}^0$ could be smaller by a factor of $\epsilon$ 
from what is expected
on the basis of U-spin symmetry with $B^+ \to \pi^+ K^0$. Therefore,
despite the relative $1/\lambda ^2$ factor \cite{Fleischerresc}, 
the overall FSI effects in $B^+ \to K^+ \bar{K}^0$
need not be much larger than those in $B^+ \to \pi ^+ K^0$.
Thus, from the smallness of FSI effects in $B^+ \to K^+\bar{K}^0$
one cannot infer that such effects are negligible elsewhere.
In fact, a $\Delta $-induced term, such as that
in $B^+ \to K^+\bar{K}^0$, 
is present in all
formulas in which the SD penguin ${P}$ contributes. This 
leads (in the SU(3) limit) to the replacement of the original
SD penguin amplitude $\bar{P}$ by the effective penguin $\tilde {P}$ given by 
Eq. (\ref{tildedef}). It is only through a combined description of all
the $B \to PP $ branching ratios (and possibly asymmetries)
 such as these attempted in this paper
(ie. not just of $B^+ \to K^+\bar{K}^0$ and $B^+ \to \pi^+ K^0$ decays)
that the effects induced by terms proportional to
 $\Delta $ can be hopefully determined.

In order to study only the most important effects, we
make now three assumptions for the fits of the next Section:\\
\indent 1) First, we put $\epsilon =0$ thus
breaking SU(3) maximally.
\\
\indent 2) Second, the
present upper bound on the value of
the $B \to K^+K^-$ branching ratio ($<0.6 \cdot 10^{-6}$) 
limits the size of $\bar{u}$
quite severely.
Thus, we assume for simplicity that $\bar{u}=0$.\\
\indent 3)
Third, with no bounds set by $B \to K^+K^-$ on 
$\bar{d}$ and
$\bar{c}$ we must treat   these parameters as free. However,
while the value of $\bar{d}$ could
be complex, one expects that
$\bar{c}$ should be real (as required by the condition of
no exotics in the $s$-channel
- see Fig. 1(c); for the Regge model the corresponding expressions  
may be found in ref.\cite{Zen1999}).
 Consequently, we will have three real parameters:
${\rm Re} ~\bar{d}$, ${\rm Im} ~\bar{d}$, 
${\rm Re} ~\bar{c}$.

With $\epsilon $ fixed, our formulas depend on six real parameters:
$|\bar{T}|$, $\bar{P}'$, $\bar{S}'$, ${\rm Re}\bar{d}$,
${\rm Im}\bar{d}$, ${\rm Re}\bar{c}$ (in addition to weak phases).
This may be compared to the approach of \cite{He2001} which is less specific as
to the origin of strong phases and involves seven independent hadronic
parameters.

\section{Fits}

In order to estimate the effects which SU(3)-breaking rescattering may induce,
we performed fits to the available branching ratios of B decays.
We decided to 
compare the case with no FSIs
(or with SU(3)-symmetric Pomeron-induced FSIs only) 
to the following two cases:\\
(a) SU(3)-breaking Pomeron-exchange-induced FSIs only,\\
(b) both Pomeron-exchange-induced and non-Pomeron SU(3)-breaking FSIs.\\
Since one of the objectives of this paper was to test the FSI effects,
we assumed that the relative strong SD phases are negligible, ie.
all direct CP violation effects involve only
the long-distance strong phases
generated by the FSI term $i {\bf \Delta W_{inel}}$
 in Eq.(\ref{fullFSI}).

As the data constraining 
our fits we used only 
the branching ratios of the $B \to PP$ decays
(ie. we did not include the data on asymmetries).
The first and the second column of Table 3 specify 
the decay channels $i$ considered  and the values of the experimental
branching ratios (and their errors) as used in our paper
(from \cite{CR2001,PDG,ICHEP2002}). In the calculations
themselves, the branching ratios were corrected for the deviation
of the ratio of the $\tau _{B^+}$ and $\tau _{B_0}$ lifetimes from unity.
The sum 
over all these decay channels $i$
of the deviations between the
experimental and theoretical branching ratios ${\bf B}_i$
normalized to
their experimental errors:
\begin{equation}
\label{tominim}
f(SD~ampl;~FSI~param.) = \sum _i \frac{({\bf B}_i(theor)-{\bf B}_i(exp))^2}{(\Delta 
{\bf B}_i(exp))^2},
\end{equation}
was subject to the
minimization procedure (see eg.\cite{perturb,HLLD}). Note that 
in our fits we used not only
the $B \to \pi \pi$ and $B \to \pi K$ branching ratios
(as in \cite{perturb}), but also
the remaining $B \to PP $ branching ratios not considered  elsewhere 
(in particular, those for $B \to K \eta , K \eta '$).
We performed several different fits, first keeping some of the
arguments of $f$ in Eq.(\ref{tominim}) fixed, and then letting them free.
The minimization procedure gave the best values of
$\bar{T}$, $\bar{P}'$, and $\bar{S}'$ (for different values 
of weak phase $\gamma $) as well as the values of the FSI parameters.
The fits permitted predictions of
CP asymmetries in $B \to K\pi $, 
the values of parameters $S_{\pi \pi}$ and $C_{\pi \pi}$  describing the
behaviour of the time-dependent rates in $B^o_d(t) \to \pi ^+ \pi ^-$, etc.
Below we discuss our results in more detail.

\subsection{Pomeron-induced rescattering}
Consider first the situation with Pomeron-induced FSIs only, ie. 
$\bar{d}=\bar{c}=\bar{u}=0$.
Two cases differing with respect to the sign
of (real) $P'$ may be distinguished. 
A negative value of $P'$ 
corresponds to vanishing differences of SD strong phases 
(eg. $\delta _{P'}$, $\delta
_{T'}$),
while its positive value corresponds to
this difference being $180^o.$
Using $P'$, $|T|$, and $S'$ as free parameters, we minimized $f$ for different
values of $\gamma $ for the  no-FSI case (all $a$'s vanish), and for case (a)
above.
Dependence of the minimum value of $f$ on the value of $\gamma $ is
shown in Fig.2. 
From Fig.2 (and Table 3)
one can see that the introduction of  SU(3)-breaking Pomeron-induced
FSIs does not lead to a significant improvement
in the description of data. 
Since non-Pomeron contributions to elastic rescattering
cannot be large at $B$ mass, this result 
 is in contradiction with a recent paper \cite{Chua} which
claimed that data provide evidence 
for a large effect due to SU(3) breaking in elastic rescattering.

The preferred values of $\gamma $ are in the range of around
$85^0 < \gamma < 125^o$ ($ 0^o <\gamma < 60^o $)
for $P'<0$ ($P'>0$).
The best fit is obtained for $P'<0$ with
 $\gamma \approx 102^o $ (see Table 3),
in agreement with earlier determinations preferring $\gamma 
~\raisebox{-0.9ex}{$\stackrel{\textstyle{>}}{\sim}$}~
 90^o$
\cite{perturb,CKM2002,MF2002}.
Such a large value of $\gamma $ is in disagreement with the estimates
in the Standard Model, which lead to 
$\gamma _{SM}\approx 64.5^o \pm 7^o$ \cite{CKM2002}, or, more conservatively,
to the region  of $50^o < \gamma < 80^o$
(see eg. \cite{Fleischer2002,BurasParodi2002,Ciuchini2001}).
The approach of ref.\cite{He2001} permits slightly smaller values of $\gamma $,
in the range of approximately $75^o-85^o$, at the cost of introducing seven
independent parameters in place of $|T|$, $P'$, and $S'$ (see also the 
next subsection).
Table 3 shows that 
the inclusion of SU(3)-breaking Pomeron-induced FSIs enhances 
the value of the $S'/P'$ ratio when extracting it from data.

\subsection{Inelastic rescattering}

Since even when $\bar{u}=\epsilon = 0$ there are still three real FSI
parameters (${\rm Re} ~\bar{d}$, ${\rm Im} ~\bar{d}$, 
${\rm Re} ~\bar{c} $), it is
instructive to consider
first the two limiting cases when 1) $|\bar{d}| \ll |\bar{c}|$ and 
2) $|\bar{d}| \gg |\bar{c}|$.
In order to study these cases, 
we assume $\bar{d}=0$ or $\bar{c}=0$ respectively.
The results of our fits for the $P'>0$ $(P'<0)$ cases 
are shown in Fig.3a (Fig.3b).
Solid (dashed) lines correspond to $\bar{d}=0$ ($\bar{c}=0$).

Clarification of how the curves in Fig.3 were obtained is in order.
The approximation leading to Eq.(\ref{fullFSI})
was based on the assumption that FSIs may be treated
perturbatively.
Consequently, the FSI parameters $\bar{d}$, $\bar{c}$ 
cannot be too large.
Consider for example the $d T$ correction to the penguin SD
amplitude $P$.
Since the ratio of $|P|/|T|$ is expected to be around
$0.3$ (in our fits without FSIs we have $=0.73/2.58= 0.28$),
the admissible value of $|\bar{d}|$ should be smaller than that
number. 
Consequently, in the analysis leading to Fig.3
we limited the region of parameter values
to $|{\rm Re}~\bar{d}|<0.25$, $|{\rm Im}~\bar{d}|<0.25$.
In order to give a feeling for the expected
scale of FSI parameters, let us recall that the contribution to $|u_+|$
arising from
quasi-elastic non-Pomeron rescattering is fully calculable in the Regge model,
and in ref.\cite{LZ2002} it was estimated to be of the order of
$0.04-0.05$ . The value of $|\bar{d}|$ of the order of $0.1$ or $0.2$ 
could therefore represent the sum of
contributions from several 
intermediate channels while being still acceptable as corresponding to a
perturbative realm.

When our restrictions on the allowed values of  
$|{\rm Re}~\bar{d}|, |{\rm Im}~\bar{d}|$
are relaxed, the global minima seen in Fig.3
are still present with the same values of $f$.
For the $P'>0$ case (Fig.3a), the relevant curve lies only 
slightly below that
shown.
For the $P'<0$ case (Fig.3b), 
the minimum of the dashed curve on the right (at $\gamma \approx 130^o$)
becomes deeper with the value of $f$ comparable to its value
at $\gamma \approx 60^o$. However, the corresponding value of $|\bar{d}|$
becomes significantly larger than $0.25$.
The fitted values of $|\bar{c}|$ are of the order of 0.25 also
when $|\bar{c}|$ is not restricted. In the presented fits
no restrictions on $\bar{c}$ were imposed.

Comparison with Fig.2 shows that the minima of $f$ treated 
as a function of $\gamma $ are now deeper and
significantly shifted when compared with the no-FSI case.

For the $P'>0$ case,  we have:
in the $\bar{d}=0$ case the minimum of $f(\gamma )$ 
appears at $\gamma \approx
50^o$ with a value of $f$ at minimum being $12.2$ and $\bar{c} = -0.28$
 (Fig.3a, solid line),
while in the $\bar{c}=0$ case $\gamma \approx 80^o$ is singled out
with $f =13.3$ and ${\rm Re}~\bar{d}=+0.25$, ${\rm Im}~\bar{d}=-0.21$
(Fig.3a, dashed line). 
The reduced $\chi ^2_{red} = f/(N-k)$, with $N=15$ used as the number of data
points, and $k$ being the number of independent parameters, goes down from
$\chi ^2_{red}$ around $25/(15-4) \approx 2.2 $ for the no-FSI case to
$\chi ^2 _{red}$ around $ 1.2 - 1.4$ when FSI is taken into account.

For the $P'<0$ case, the minima of $f(\gamma )$ are
significantly deeper:
 in the $\bar{d}=0$ case there is a slight shift in $\gamma $
(from around $102^o$ to around $90^o$ ) with the value of 
$f(\gamma )$ at minimum being $8.84 $ and $\bar{c}=0.24$ (Fig.3b, solid line); 
in the $\bar{c}=0$ case the shift in $\gamma $ is larger
and a minimum appears  at $\gamma = 57^o$
with the value of $f(\gamma )$ at minimum being $7.61$ (Fig.3b, dashed line). 
In the latter case the fitted values of FSI parameters 
are
\begin{eqnarray}
{\rm Re} ~\bar{d}& \approx & -0.22\\
{\rm Im} ~\bar{d}& \approx & +0.21
\end{eqnarray}
In both cases the value of $\chi ^2_{red}$ is about $0.9$.
The second minimum of the dashed line in Fig.3b
at $\gamma \approx 130^o$ corresponds to a different
sign of ${\rm Re}~ \bar{d}$. 
When the restriction on the size of $|\bar{d}|$ is relaxed, this minimum
becomes as deep as that at $\gamma = 57^o$. 
Then, however, the value of
 $|\bar{d}|$ is
much larger than 0.25.
Since $\chi ^2_{red}$ is significantly smaller for $P'<0$, 
we restrict further discussion to this case. 

In Fig.4, relaxing for a moment the
assumption $\epsilon = 0$, we show the $\epsilon$-dependence of the minimal
values of $f$ for $P'<0$ and for fixed values of $\gamma $ 
in the two cases
of $\bar{d}=0$ (Fig.4a) and $\bar{c}=0$ (Fig.4b). 
 The region of small $\epsilon $
seems to be preferred in both cases.
In this analysis, as in that leading to Fig.3, 
the values of  $\bar{d}$  were 
restricted to $|{\rm Re}~\bar{d}|<0.25$, $|{\rm Im}~\bar{d}|<0.25$,
while the values of $\bar{c}$ were set free.

In the most general fit (with $P' <0$), we assumed $\epsilon =0$ 
and simultaneously  treated all three FSI parameters
(${\rm Re}~ \bar{d}$, ${\rm Im}~\bar{d}$, $\bar{c}$) as free.
In Fig.5a we show the contour plot of the minimum of $f$ 
treated as a function
of complex $\bar{d}$.
The fitted values of $\bar{c}$ are not shown but in the region around 
${\rm Re}~ \bar{d}=-0.22$, ${\rm Im}~ \bar{d}=+0.21$ (point
$X$)
they turn out to be close to $0$.
Thus, allowing $\bar{c}$ to be free does not lead far away 
from the minimum found
before for the $\bar{c}=0$ case.The corresponding $\chi ^2_{red}$ is 
around $1.0$.
The fitted values of $|\bar{c}|$ turn out to be smaller than $0.25$
for all of $\bar{d}$ in Fig.5a with the exception of a thin slice
on the right (for ${\rm Re}~\bar{d} > 0.20$ and 
${\rm Im} ~\bar{d} < 0.05 $).

In order to show what happens 
for other negative as well as for positive ${\rm Re}~\bar{d}$, below we 
present also fits performed at two additional
points ($p1$) and ($p2$):

\begin{eqnarray}
{{\rm point} ~ p1:}&&{\rm Re}~\bar{d}=-0.10\nonumber \\
&&{\rm Im}~\bar{d}=+0.15
\end{eqnarray}
and
\begin{eqnarray}
{{\rm point} ~ p2:}&&{\rm Re}~\bar{d}=+0.15\nonumber \\
&&{\rm Im}~\bar{d}=+0.15
\end{eqnarray}
The $B \to PP$ branching ratios corresponding 
 to the four cases ($\bar{d}=0$, $\bar{c}=0$, point ($p1$), point ($p2$)) are
gathered in Table 3 together with other fit details.

As can be seen from Table 3, the quality of the description of branching ratios
at 
points ($p1$), ($p2$) is essentially the same as that at minimum (point $X$, 
$\bar{c} \approx 0$). 
Table 3 shows also that the dominant contribution to $f$ comes from
the $2\sigma $ discrepancy between the experimental and the fitted
$B^0_d \to \pi ^0 K^0$ branching ratios
(a similar problem with this decay
channel can be observed in other papers, see eg. \cite{Chua} ). 
In a recent paper \cite{GroRos0307095},
the question of a potential discrepancy in the sum rule relating 
the branching ratios in $B^{+},B^0_d \to K \pi$
decays was discussed and it was suggested that the experiment hints at
a slight enhancement of both modes involving $\pi ^0$.
In our fits (as in \cite{Chua}), however,
the measured branching ratio of $B^+ \to \pi ^0 K^+$ is well described.

Fig.5b gives the contour plot of the corresponding 
fitted values of $\gamma $. In the region around 
points $X$ and $p1$ the fitted values of $\gamma $
seem to be in agreement with the conservative SM expectation of
$50^o < \gamma _{SM} < 80^o $, so this part of
the complex $\bar{d}$ plane may be called the "SM" region.

\subsection{CP asymmetries}
With the values of the FSI (and other) parameters fixed, one can
attempt the calculation of CP-violating observables.
The CP-violating asymmetries in $B \to K \pi $ decays defined as
\begin{equation}
\label{KpiCPasym}
A_{CP}(B \to K \pi)=
\frac{\Gamma (\bar{B}\to\bar{K} \pi)-\Gamma (B \to K \pi)}
{\Gamma (\bar{B}\to\bar{K} \pi)+\Gamma (B \to K \pi)}
\end{equation}
($B=B^0_d,B^+$, $\bar{B}=\bar{B}^0_d,B^-$)
were calculated for all four cases
under discussion.  The relevant predictions
are given in Table 4 together with the 
experimental data
(\cite{ICHEP2002,CLEO,Belle2002,BaBar2002,Tomura,Belle0304035,BaBar0303028}) 
as averaged in
\cite{CGR0306021}.
The "SM" region of small $\bar{c}$ 
and negative ${\rm Re}~\bar{d}$ 
(represented by points $X$ ($\bar{c}=0$) and $p1$
($\bar{c}=-0.11$)) seems
to describe the experimental $B \to K \pi$ CP
asymmetries somewhat better than the $\bar{d}=0$
case or the region of positive ${\rm Re}~\bar{d}$ (ie. point $p2$) do: 
our FSI approach prefers negative $B^0 \to \pi ^- K^+$ asymmetry,
in agreement with the experiment and in disagreement with 
the predictions of ref.\cite{perturb}. Although 
the $B \to K\pi $ asymmetries are experimentally small, 
they might provide important model
tests (see eg. \cite{GerardSmith2003}).

In view of the recent BaBar measurement \cite{BaBar2003a} favouring a large 
negative asymmetry in
$B^{+} \to \pi ^{+} \eta$ decays, we have computed the asymmetries in all
$B^+ \to \pi^+ \eta (\eta ')$ and $B^+ \to K^+ \eta (\eta ')$ decays.
The results are given in Table 5 
with the data \cite{BaBar2003a,BaBar0303046,Belle0207033} 
averaged as in \cite{CGR0306021}. Contrary to
the BaBar result,
our $B^+ \to \pi ^+ \eta$ asymmetry is small and positive for
$\gamma $ in the "SM" region. 
On the other hand, our $K^+ \eta $ asymmetry 
(fairly large when compared with other asymmetries)
seems to agree 
with the data.
Problems with the simultaneous description of $\pi ^+ \eta$ and
$K^+ \eta $ asymmetries have been noted in \cite{CGR0306021} as well.

We have also calculated parameters relevant
for the time-dependent rates in $B^0_d(t) \to \pi^+ \pi^-$ \cite{Gronau89}, 
ie.:
\begin{equation} 
S_{\pi \pi}=\frac{2~{\rm Im} \lambda _{\pi \pi}}{1+|\lambda _{\pi \pi}|^2}
\end{equation}
 and 
 \begin{equation}
 C_{\pi \pi}=\frac{1-|\lambda _{\pi \pi}|^2}{1+|\lambda _{\pi \pi}|^2},
 \end{equation}
 where
 \begin{equation}
 \lambda _{\pi \pi} \equiv e^{-2 i \beta} 
 \frac{A(\bar{B}^0_d \to \pi ^+ \pi^-)}{A(B^0_d \to \pi ^+ \pi ^-)}.
 \end{equation}
Our predictions are given in Table 6.
Although the experimental results from Belle and BaBar
\cite{BaBar0207055,Belle0301032}
 still exhibit the well-known discrepancies \cite{Belle2003,BaBar2003},
the "SM" region of small (negative) $\bar{c}$ and 
negative ${\rm Re}~\bar{d}$ (with the value of $\gamma$ 
close to the SM expectations) seems favoured again.

For the time-dependent rates in $B^0_d(t) \to \eta ' K_S $, the effect of
final-state interactions is negligible. 
Indeed, the relevant amplitudes are dominated by the
$\bar{P}'$ and $\bar{S}'$ amplitudes 
(in particular, the FSI correction is dominated by terms
proportional to $\bar{P}'$, see Table 2). 
Thus, all important terms  have the same
 weak phase. Consequently, one obtains
$S_{\eta ' K_S} \approx \sin 2\beta $, $C_{\eta ' K_S} \approx 0 $,
in agreement with the experimental averages (from \cite{CGR0306021}) of 
$S_{\eta ' K_S} =+0.33 \pm 0.25 $, 
$C_{\eta ' K_S}= -0.18 \pm 0.16$ .

The $B^+ \to \pi ^+ \pi ^0$ asymmetry is predicted to be 
zero (cf. Tables 1 and 2),
in agreement with its experimental value of $-0.07 \pm 0.14$ 
(average from \cite{CGR0306021}).

Although apart from the discrepancy in sign with the most recent BaBar 
$\pi ^+ \eta$ result there seems to be a hint of agreement with other
asymmetries, one has to remember that these (and other)
 predictions for asymmetries  may be affected by 
 the inclusion of the charming penguin
contribution \cite{Ciuchini,CKM2002}.

\section{Conclusions}
In this paper we have analysed the contributions from both elastic
and inelastic SU(3)-breaking final-state interactions in $B$ decays to two
light pseudoscalar mesons ($B \to PP $).

We have found that the inclusion of an experimentally determined pattern of 
SU(3) breaking in Pomeron-induced rescattering
enhances the value of the $S'/P'$ ratio when extracting it from the fit
to the $B \to PP $ branching ratios. However,
taking this rescattering into account
 does not lead to any significant change in the overall fit. 
Since at the energy of $s = m^2_B$ 
the inclusion of non-Pomeron elastic rescattering may lead to small
corrections only, analyses incorporating full elastic FSIs can lead neither 
to a significant improvement
in the quality of data description, nor to 
the extracted value of $\gamma $ being
substantially shifted towards the SM expectation.

We have pointed out that a small value of the $B \to K^+K^-$ branching ratio 
{\em does not
imply} negligible inelastic rescattering effects in other $B \to PP$ decays.
This conclusion follows from the fact
 that rescattering in the $B \to K^+K^-$ decay 
is independent of two of the three parameters describing the totality of 
inelastic FSIs: 
one related to the $u$-loop long-distance
penguin (in a resonance channel), and the other one describing 
quark rearrangement (in an exotic channel).
As for $B^+ \to K^+\bar{K}^0$, with U-spin symmetry probably broken by
final-state interactions,
this decay was argued to be 
less helpful in the determination of the size of rescattering effects
than originally suspected. Its importance in the determination of the
size of rescattering effects (ie. the size of the $u$-loop long-distance
penguin) would then lie not just in its relation to 
$B^+ \to \pi ^+ K^0$, but, more properly, in its relation to  
all other $B \to PP$
decay channels. 

Finally, after neglecting the
relative strong phases of short-distance amplitudes, we have carried out
fits to the available $B \to PP $ branching ratios 
 with all
elastic and inelastic SU(3)-breaking rescattering effects taken into account.
The only neglected but potentially important corrections were those due
to the intermediate states composed of charmed mesons. 
Our fits show the importance of rescattering effects and weakly hint
at the value of $\gamma $ compatible with SM expectations.
However, other values of $\gamma $ are also possible.
Narrowing the range of admitted values of $\gamma $ will require
taking into account the experimental data on asymmetries in addition to those
on branching ratios. In this paper we
used the values of rescattering parameters as
determined from the fits to the branching ratios,
and predicted
several CP-violating observables (CP asymmetries in $B \to K \pi$ decays, 
$S_{\pi \pi}$ and $C_{\pi \pi}$ for the $B^0_d(t) \to \pi ^+ \pi ^-$
time-dependent decay rates etc.). Again, weak agreement with the data 
 (with the notable exception of the $B^+ \to \pi^+ \eta$ asymmetry) was found
 for $\gamma $ close to the SM expectations.

This work was supported in part 
by the Polish State Committee for Scientific
Research (KBN) as a research project in 
years 2003-2006 (grant 2 P03B 046 25).

\newpage
FIGURE CAPTIONS

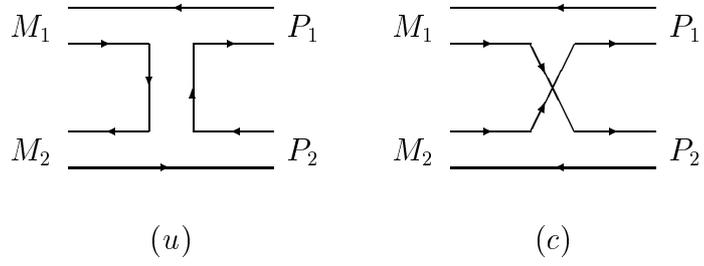
Fig. 1. Types of rescattering diagrams: (u) uncrossed, (c) crossed.

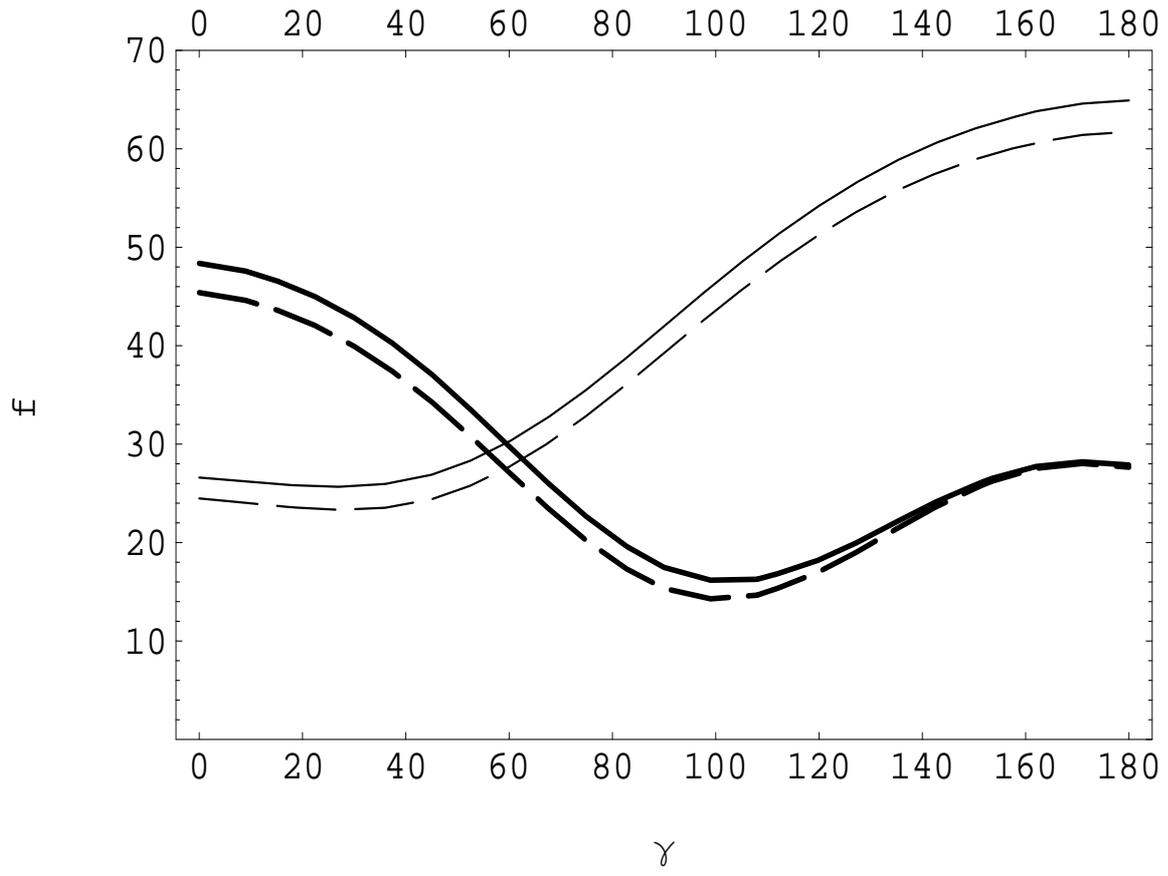
Fig. 2. Dependence of minimized function $f$ (Eq.(\ref{tominim}))
on $\gamma $: thin lines - $P'>0$, thick lines - $P'<0$;
 solid lines - no FSI/SU(3) symmetric Pomeron-induced FSI,
  dashed lines - SU(3)-breaking
Pomeron-induced FSI.

Fig. 3. Dependence of minimized function $f$ (Eq.(\ref{tominim}))
on $\gamma $ for full FSI: 
(a) $P'>0$, (b) $P'<0$; solid lines - $\bar{d}=0$, unrestricted $|\bar{c}|$; 
dashed lines - $\bar{c}=0$, $|{\rm Re}~ \bar{d}|<0.25$,
$|{\rm Im}~ \bar{d}|<0.25$.

Fig. 4. Dependence of minimized function $f$ on $\epsilon $
for full FSI and different values of $\gamma $: (a) $\bar{d}=0$, 
unrestricted $|\bar{c}|$;
(b) $\bar{c}=0$, $|{\rm Re}~\bar{d}|<0.25$, $|{\rm Im}~\bar{d}|<0.25$. 

Fig. 5. (a) Contour plot of minimized function $f$ in complex $\bar{d}$
plane. Positions of the minimum ($X$) and of the selected points $p1$, $p2$
are indicated.
(b) Contour plot of fitted values of $\gamma $ in complex $\bar{d}$ plane.

\newpage

\begin{table}[t]
\caption{Branching ratios of B decays
 (in units of $10^{-6}$) }
\label{FSIfits}

\begin{tabular}{|cc|cccccc|}
\hline 
decay & expt  
& \multicolumn{6}{c|}{$P'<0$}
 \\
&& no FSI & 
Pomeron &$\bar{d}=0$ & $\bar{c}=0$ &  p1
&  p2\\
\hline

$B^+ \to$ $ \pi ^+ \pi ^0$           & 
$5.8 \pm 1.0 $&  $4.85$ 
&$4.79$ &$5.23$ &$5.38$ &$5.54$ &$5.86$ \\

\phantom{$B^+ \to$} $K^+\bar{K}^0 $  &  
$ 0.0 \pm 2.0$&   $0.57$
&$0.51$ &$0.54$ &$1.09$ &$1.02$ &$0.87$ \\

\phantom{$B^+ \to$} $\pi ^+\eta $    &
$2.9 \pm 1.1$&  $2.13$ 
&$2.13$ &$3.47$ &$2.90$ &$2.60$ &$2.50$ \\

\phantom{$B^+ \to$} $\pi ^+\eta '$   &
$0.0 \pm 7.0 $&  $1.06$ 
&$1.03$ &$1.69$ &$1.39$ &$1.25$ &$1.22$ \\
\hline

$B^0_d \to$ $\pi ^+ \pi ^-$          &
$ 4.7 \pm 0.5 $& $4.93$  
&$4.93$ &$5.19$ &$4.77$ &$4.79$ &$4.62$ \\

\phantom{$B^0_d \to$} $\pi ^0 \pi ^0$& 
$1.9 \pm 0.7 $&  $0.55$ 
 &$0.56$ &$1.98$ &$1.85$ &$0.82$ &$1.31$\\

\phantom{$B^0_d \to$} $K^+K^-$       & 
$0.0 \pm 0.6 $ &  $0$  
&$0$ &$0$ &$0$ & $0$ &$0$\\

\phantom{$B^0_d \to$} $K^0 \bar{K}^0$&
$0.0 \pm 4.1 $ &  $0.53$  
 &$0.48$ &$0.50$ &$1.02$ &$0.95$ & $0.87$\\
\hline

$B^+ \to $ $\pi ^+ K^0$              & 
$18.1 \pm 1.7 $ & $18.28$  
&$18.51$ &$19.70$ &$19.15$ &$18.98$ &$20.53$ \\

\phantom{$B^+ \to$} $\pi ^0 K^+$     &
$12.7 \pm 1.2$ &  $12.96$ 
&$12.87$ &$12.47$ &$12.15$ &$12.34$ &$12.76$\\

\phantom{$B^+ \to$} $\eta K^+$       &
$4.1 \pm 1.1 $ &  $2.45$ 
&$3.05$ &$3.64$ &$4.18$ &$4.07$ &$4.24$ \\

\phantom{$B^+ \to$} $\eta ' K^+$     & 
$75.0 \pm 7.0 $&  $72.85$ 
&$72.09$ &$69.31$ &$69.07$ &$69.53$ &$69.60$ \\
\hline

$B^0_d \to$ $\pi ^- K^+$             & 
$18.5 \pm 1.0 $&  $18.90$  
&$18.90$ &$17.57$ &$18.89$ &$18.99$ &$18.10$ \\

\phantom{$B^0_d \to$} $\pi ^0 K^0$   &
$10.2 \pm 1.5 $&  $6.38$ 
&$6.53$ &$6.79$ &$7.16$ &$7.04$ &$7.37$ \\

\phantom{$B^0_d \to$} $\eta K^0$     &
$0.0 \pm 9.3  $&  $1.83$ 
&$2.43$ &$4.28$ &$2.50$ &$2.29$ &$5.36$ \\

\phantom{$B^0_d \to$} $\eta ' K^0$   &
$56.0 \pm 9.0$ &  $67.07$ 
&$66.62$ &$65.68$ &$66.51$ &$65.37$ &$65.06$\\
\hline
$f_{min}$ &&16.05&14.25&8.84&7.61&9.70&8.86\\
$|\bar{T}|$&&2.58&2.56&2.41&2.71&2.69&2.66\\
$\bar{P'}$&&-4.14&-4.24&-4.34&-6.17&-5.98&-5.53\\
$\bar{S'}$&&-1.77&-2.27&-2.09&-1.53&-1.41&-1.52\\
$\gamma _{fit} $&
&$103^o$&$101^o$&$89^o$&$57^o$&$78^o$ &$99^o$\\
\hline
$\bar{c}$&&&&$+0.24$&$0$&$-0.11$&$+0.18$\\
${\rm Re}~\bar{d}$&&&&$0$&$-0.22$&$-0.10$&$+0.15$\\
${\rm Im}~\bar{d}$&&&&$0$&$+0.21$&$+0.15$&$+0.15$\\
\hline
\end{tabular}\phantom{xx}

\end{table}

\newpage
\begin{table}[t]
\caption{Asymmetries in $B\to K\pi $ decays}
\label{FSIfitsa}
\begin{center}

\begin{tabular}{|cc|cccc|}
\hline 
decay & expt 
& \multicolumn{4}{c|}{$P'<0$}

\\
& &  $\bar{d}=0$ 
& $\bar{c}=0$ & p1 &  p2   \\
\hline

$B^+ \to$ $ \pi ^+ K ^0$           & 
$-0.032 \pm 0.066 $  & $0$ &$+0.09$&   $+0.05 $ &$-0.07$ \\
$B^+ \to$ $ \pi ^0 K ^+$           & 
$+0.035 \pm 0.071 $  
&$-0.04$ &$-0.10$ & $-0.03$ &$+0.03$ \\
$B^0 \to$ $ \pi ^- K ^+$           & 
$-0.088 \pm 0.040 $& $+0.03$ 
&$-0.10$ &   $-0.07 $ &$+0.08$ \\
$B^0 \to$ $ \pi ^0 K ^0$           & 
$0.03 \pm 0.37 $  & $+0.07$ 
&$+0.13$&   $+0.04 $ &$-0.05$ \\

\hline
\end{tabular}

\end{center}
\end{table}

\newpage
\begin{table}[t]
\caption{Asymmetries in $B^+ \to \pi ^+ \eta (\eta ') $ and
$B^+ \to K^+ \eta (\eta ') $ decays}
\label{FSIfitsb}
\begin{center}

\begin{tabular}{|cc|cccc|}
\hline 
decay & expt 
& \multicolumn{4}{c|}{$P'<0$}

\\
&&    $\bar{d}=0$ 
& $\bar{c}=0$ & p1 &  p2   \\
\hline

$B^+ \to$ $ \pi ^+ \eta$           & 
$-0.51 \pm 0.19 $  &$0$ &$+0.10$&   $+0.06 $ &$-0.09$ \\
$B^+ \to$ $ \pi ^+ \eta '$           & 
&$0$ &$+0.10$ & $+0.06$ &$-0.10$ \\
$B^+ \to$ $ K ^+ \eta$           & 
$-0.32 \pm 0.20 $& $+0.23$ 
&$-0.39$ &   $-0.49 $ &$+0.32$ \\
$B^+ \to$ $ K ^+ \eta '$           & 
$-0.002 \pm 0.040 $  & $-0.01$ 
&$+0.01$&   $+0.01 $ &$-0.01$ \\

\hline
\end{tabular}

\end{center}
\end{table}

\newpage
\begin{table}
\caption{CP-violating parameters in time-dependent rates for 
$B\to \pi^+ \pi^- $}
\label{c}
\begin{center}

\begin{tabular}{|cc|cccc|}
\hline 
parameter & experiment  
& \multicolumn{4}{c|}{$P'<0$}
\\
&Belle   &  $\bar{d}=0$ 
& $\bar{c}=0$ & p1 &  p2   \\
&BaBar&&&&\\
\hline

$S_{\pi \pi}$           & 
$-1.23 \pm 0.41^{+0.08}_{-0.07}$  
 &$-0.12$ &$-0.78$&   $-0.23 $ &$+0.49$ \\
&$-0.40\pm 0.22\pm 0.33 $ &&&&\\
$C_{\pi \pi}$           & 
$-0.77\pm 0.27 \pm 0.08 $  
 &$-0.05$ &$-0.21$&   $-0.08 $ &$+0.11$ \\
&$-0.19 \pm 0.19 \pm 0.05 $ &&&&\\
\hline
\end{tabular}

\end{center}
\end{table}

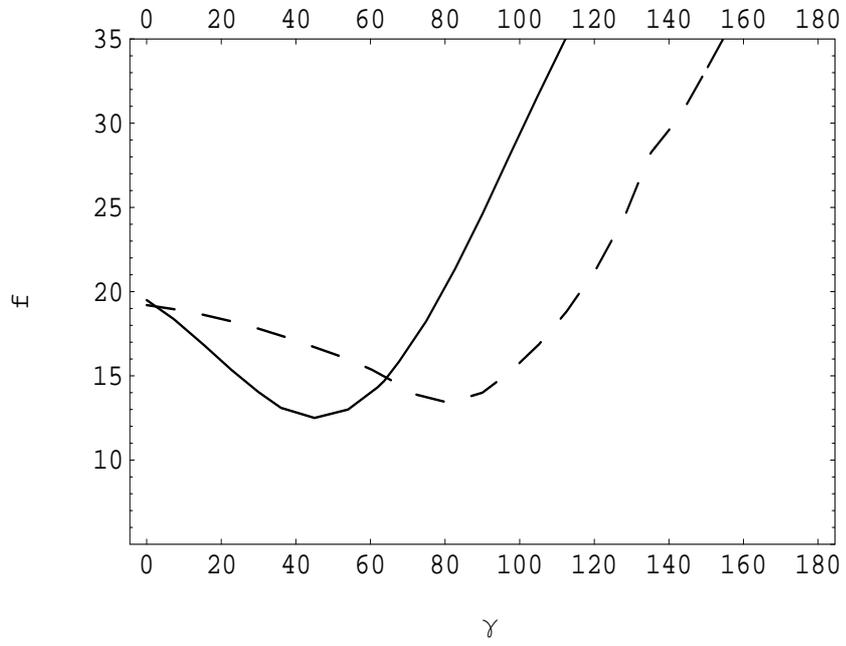
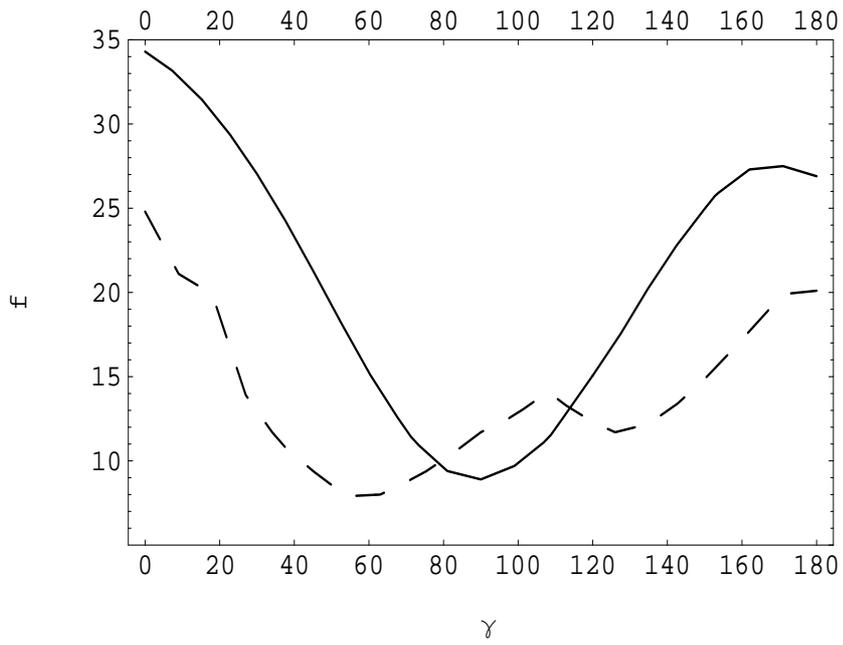
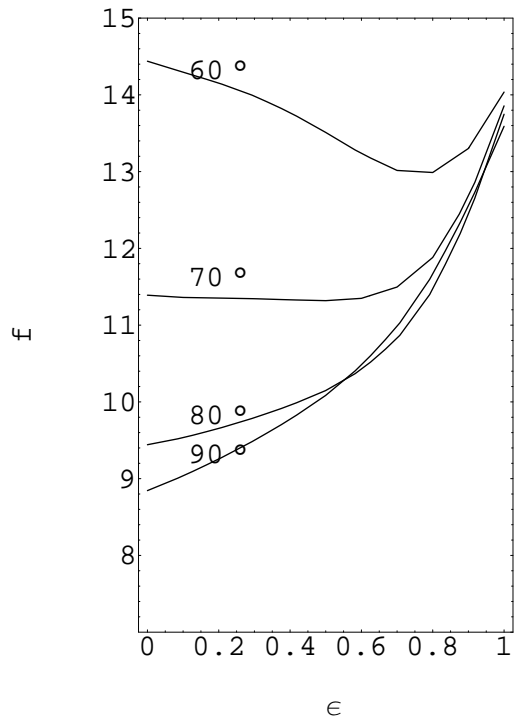
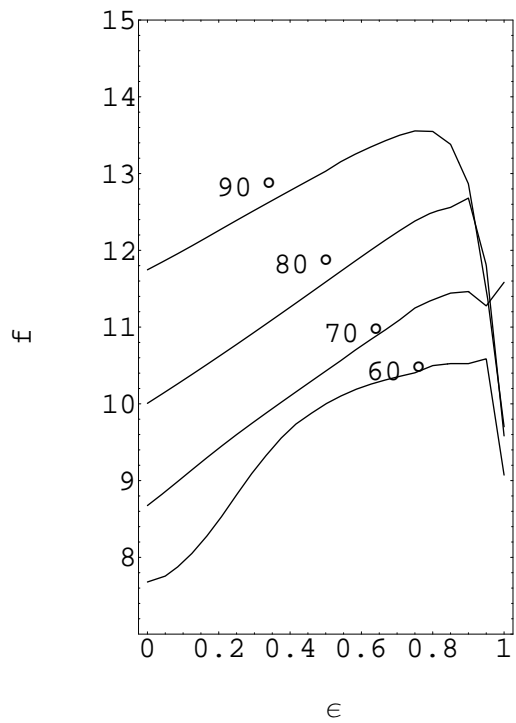
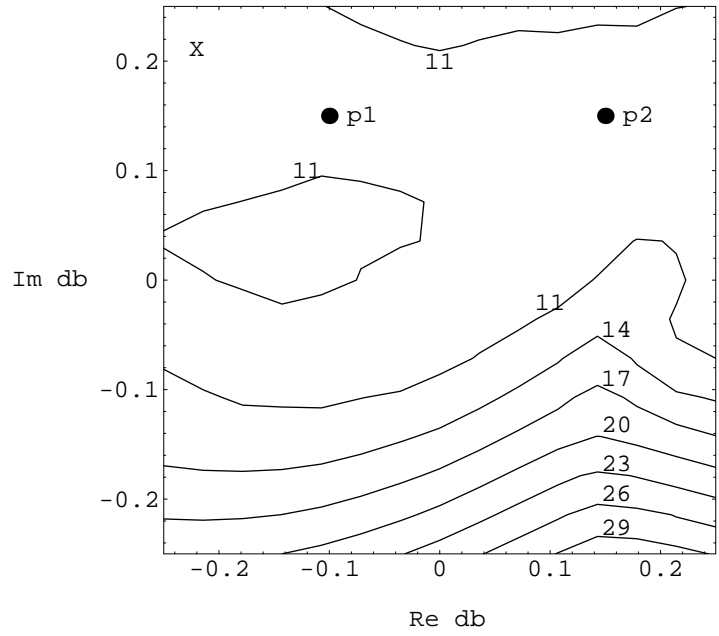
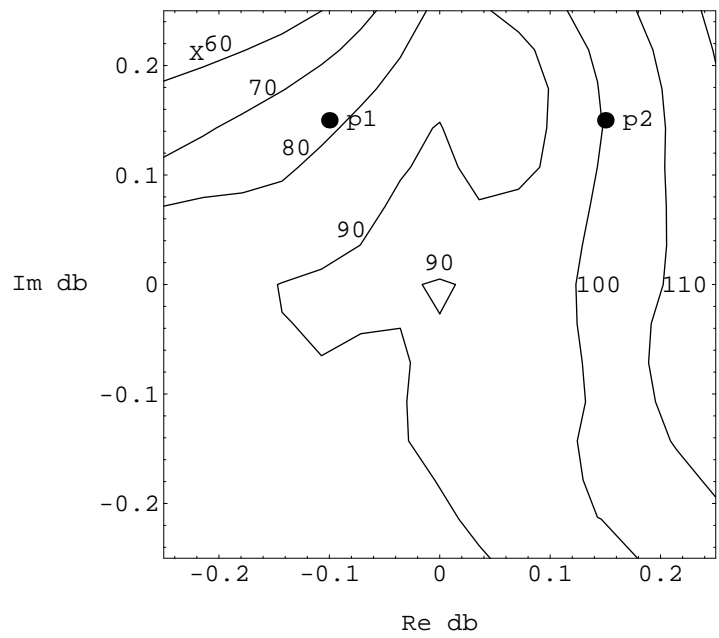
\end{document}